\documentclass{mc2005}

\title{Modeling heavy ion ionization loss in the MARS15 code}

\author{
{\bf I.L. Rakhno\thanks{Corresponding author}, N.V. Mokhov, and S.I. Striganov}\\
Fermi National Accelerator Laboratory\\
MS 220, Batavia, Illinois 60510-0500\\
rakhno@fnal.gov; mokhov@fnal.gov; strigano@fnal.gov\\[12pt]  }
\newcommand{\authorshort}{I.L. Rakhno, N.V. Mokhov, S.I. Striganov}

\newcommand{\titleshort}{Modeling heavy ion ionization loss in the MARS15 code}

\hypersetup{
pdfauthor   = {\authorshort},
pdftitle    = {\titleshort},
pdfsubject  = {Monte Carlo 2005},
pdfkeywords = {Use keyword list from abstract}}

\usepackage{epsfig}
\begin{document}

\setcounter{page}{1}
\pagestyle{plain}
\begin{figure}[htb!]
\hspace{-2cm}
\vspace{-1cm}
\epsfig{file=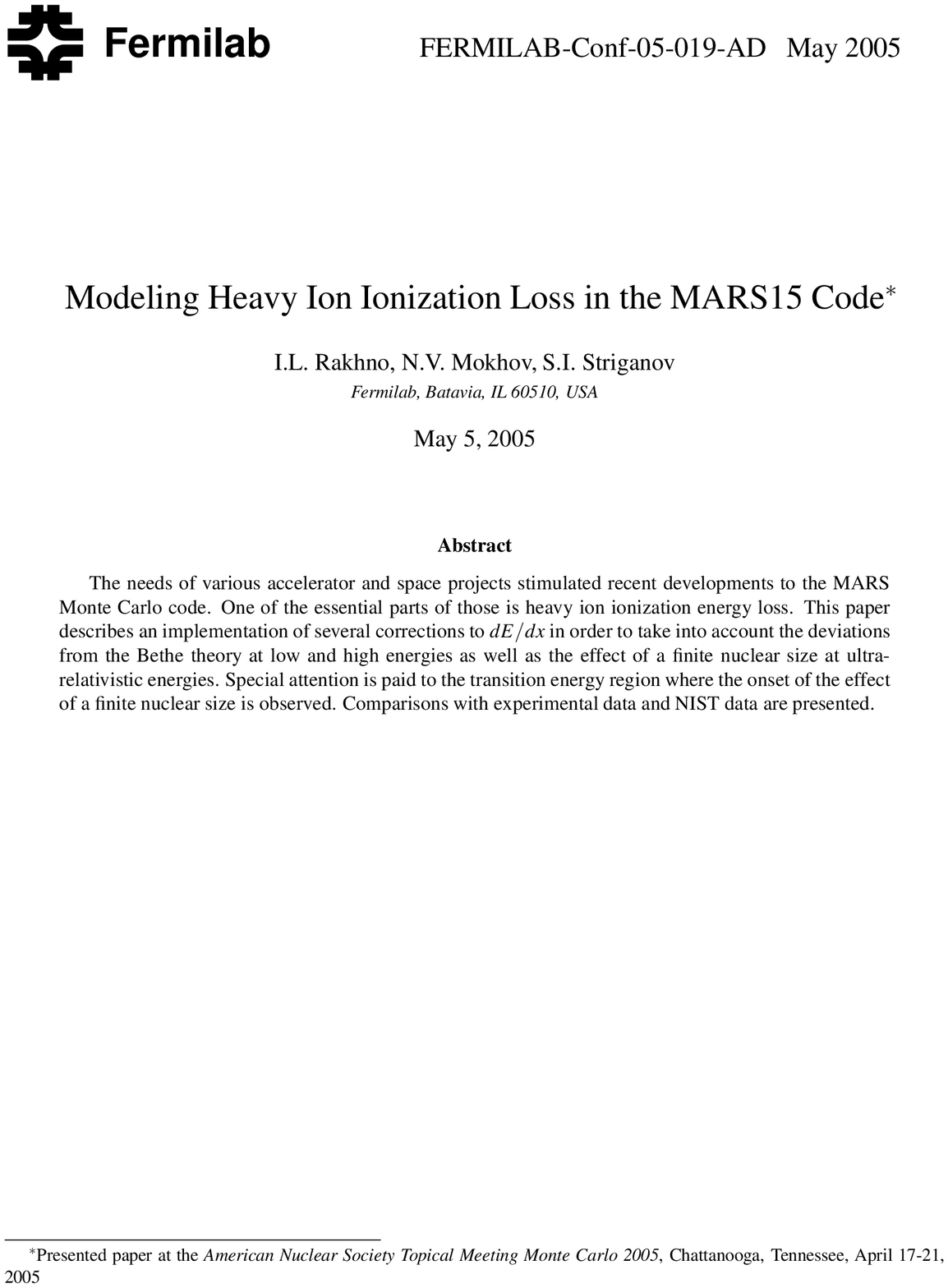,width=1.27\linewidth}
% \centering\epsfig{file=title_mc2005.ps}
\end{figure}
\clearpage

%%%%%%%%%%%%%%%%%%%%%%%%%%%%%%%%%%%%%%%%%%%%%%%%%%%%%%%%%%%%%
%%%
%%%   Leave the \maketitle and \thispagestyle{empty} commands as is, and
%%%   they will properly format the first page of the paper.
%%%
%%%%%%%%%%%%%%%%%%%%%%%%%%%%%%%%%%%%%%%%%%%%%%%%%%%%%%%%%%%%%

\maketitle
%% \thispagestyle{empty}

%%%%%%%%%%%%%%%%%%%%%%%%%%%%%%%%%%%%%%%%%%%%%%%%%%%%%%%%%%%%%
%%%
%%%   Use the abstract environment to define the abstract.  It will be
%%%   properly sized and placed on the page.  Leave a \vspace{1em}
%%%   in front of the Key Words list, and provide no more than five
%%%   key words.
%%%
%%%%%%%%%%%%%%%%%%%%%%%%%%%%%%%%%%%%%%%%%%%%%%%%%%%%%%%%%%%%%

\begin{abstract}
The needs of various accelerator and space projects stimulated recent 
developments to the MARS Monte Carlo 
code. One of the essential parts of those is heavy ion ionization energy loss. 
This paper describes an implementation of several
corrections to $dE/dx$ in order to take into account the deviations from 
the Bethe theory at low and high energies
as well as the effect of a finite nuclear size at
ultra-relativistic energies. Special attention is paid to the
transition energy region where the onset of the effect of a finite
nuclear size is observed. Comparisons with experimental data and
NIST data are presented.

\vspace{1em}{\em Key Words:} Heavy ions, ionization loss, MARS15 code

\end{abstract}

%%%%%%%%%%%%%%%%%%%%%%%%%%%%%%%%%%%%%%%%%%%%%%%%%%%%%%%%%%%%%
%%%
%%%   Use section, subsection, and subsubsection commands to define the
%%%   document structure and formatting.  Enter only the text, the formatting
%%%   will be handled automatically, except that you will need to capitalize
%%%   significant words for section and subsection headings, but
%%%   capitalize only the first word for subsubsection headings.
%%%
%%%%%%%%%%%%%%%%%%%%%%%%%%%%%%%%%%%%%%%%%%%%%%%%%%%%%%%%%%%%%

\section{Introduction}

The \textsc{MARS} code~\cite{MARS} is developed for detailed Monte
Carlo modeling of hadronic and electromagnetic cascades in
realistic geometry for various accelerator, shielding, detector and space
applications.  The recent needs of the Rare Isotope Accelerator,
Relativistic Heavy-Ion Collider, Large Hadron Colider, and NASA
projects was a stimulus to implement heavy-ion collision and
transport physics into the \textsc{MARS15} code~\cite{MARS15}.
The present paper describes in detail the ionization energy loss
formalism employed in the code along with comparisons to
experimental data and some recommended data.  
Radiative energy loss of heavy ions---bremsstrahlung and $e^{+}e^{-}$
pair production---is described elsewhere. The ionization loss is of
importance for correct prediction of radiation-induced effects, \emph{e.g.}
single-event upsets, in microelectronic devices.
The lower energy limit in our stopping power model is equal to 1 keV 
per nucleon.

\section{Formalism of Ionization Loss Theory}

In our model we distinguish three energy regions.  Below 1 MeV per
nucleon and above 10 MeV per nucleon the tabulated data on proton
total stopping power from Ref.~\cite{NIST} and the Bethe
formalism, respectively, are used in combination with all the
corrections described below. Between the two energies, a
\emph{mix-and-match} procedure is used to perform an interpolation
between the approaches.  It should also be noted that the 10-MeV
limit is identical to the one used when considering the ion
effective charge (see below) and should be adjusted for some
target nuclei to get better appearance of the ionization loss
distributions.

\subsection{Bethe Theory}

% A logical division of your paper into sections makes it much easier to
% understand. Using the subsection command for subsection headings will
% correctly number and justify your subsection headings.  If you use the
% styles provided widow/orphan lines will be avoided.

The mean ionization energy loss of charged particles heavier than electrons is given
by the Bethe expression~\cite{Bethe}
\begin{equation}
\label{Bethe}
  - \frac{1}{\rho} \frac{dE}{dx} = 4 \pi N_A {r_e}\!\!^2 m_e c^2 z^2 \frac{Z}{A} \frac{1}{\beta^2} L(\beta)
\end{equation}
where $A$ and $Z$ are the target atomic mass and number, respectively, and the other variables 
have their usual meaning.  
The ionization logarithm, $L(\beta)$, is presented in the following form:
\begin{equation}
\label{Lbeta}
  L(\beta) = L_0 (\beta) + \sum_{i} {\Delta}L_i
\end{equation}
\begin{equation}
\label{Lbeta0}
  L_0(\beta) = \ln \left ( \frac{2 m_e c^2 {\beta}^2 {\gamma}^2 }{I} \right ) - {\beta}^2 - \frac{\delta}{2}
\end{equation}
where $I$ and $\delta$ are the mean excitation energy and density correction, respectively. 
When neglecting all the corrections ${\Delta}L_i$ and dealing only with the $L_0(\beta)$, the expression (\ref{Bethe})
is referred to as the Bethe equation.  The corrections ${\Delta}L_i$ described below are to take into account
the deviations from the Bethe theory for ions at both low and high energies.

\subsection{Lindhard-S{\o}rensen Correction}

% Subsection titles should start flush left, and are numbered as
% illustrated above.

% Equations should be centered and sequentially numbered to the
% flush right of the formula.
% \begin{equation}
% T_{\gamma} = 2 \pi \langle \Gamma_{\gamma} \rangle / D_{J}
% \end{equation}
% The continuation of a paragraph after an equation is not indented
% (don't leave a blank line and start a new paragraph). All paragraphs,
% as well as section or subsection headings, are separated from the
% following text by 6 pts as set by the \LaTeX~class file or the Word
% template.

Lindhard and S{\o}rensen derived a relativistic expression for
electronic stopping power of heavy ions taking into account a finite
nuclear size~\cite{LS}.  They used the exact solution to the Dirac
equation with spherically symmetric potential which describes
scattering of a free electron by an ion.  Thus, their expression,
${\Delta}L_{LS}$, provides for the corrections of order higher
than $z^2$ to ionization loss of heavy ions in both low and high
energy regimes.  At high energies the Lindhard-S{\o}rensen ($LS$)
correction replaces the previously developed Mott correction and
relativistic Bloch-Ahlen one, while at low energies
${\Delta}L_{LS}$ reduces to the Bloch non-relativistic
correction~\cite{LBNL}.

At moderately relativistic energies (see below) the following
expression derived for point-like ions is valid:
\[
  {\Delta}L_{LS} = \sum_{k=1}^{\infty} \left [ \frac{k}{\eta^2} \,\frac{k-1}{2k-1} \sin^2 (\delta_k - \delta_{k-1}) \right.
\]
\[
   \qquad \qquad \quad \, + \: \frac{k}{\eta^2} \,\frac{k+1}{2k+1} \sin^2 (\delta_{-k} - \delta_{-k-1})
\]
\begin{equation}
\label{LS}
   \qquad \qquad \qquad \left. + \frac{k}{4k^2 - 1} \:\frac{1}{\gamma^2 k^2 + \eta^2} - \frac{1}{k} \right ] + \frac{\beta^2}{2}
\end{equation}
where $\eta = \alpha z / \beta$, $\delta_k$ is a relativistic
Coulomb phase shift expressed with the argument of the complex Gamma
function (for details see Ref.~\cite{LBNL}), 
and $k$ is a parameter used in the summation over partial waves.  
At higher energies, when $\gamma m_e c
R\simeq\hbar / 2$ where $R$ is the ion radius, a modification to
the Coulomb phase shifts due to a finite nuclear size is not
negligible and the expression for ${\Delta}L_{LS}$ gets more
complicated from computational standpoint. At ultra-relativistic
energies, when $\gamma m_e c R \gg \hbar / 2$, an asymptotic
expression for $L(\beta)$ is valid.
\begin{equation}
\label{LS_ultra}
   L_{ultra} = L_0(\beta) + {\Delta}L_{LS} = \ln \left ( \frac{2 c}{R \omega_p} \right ) - 0.2
\end{equation}
where $\omega_p$ is the plasma frequency, $\sqrt{4\pi n
e^2/m_e}$\,, and $n$ is the average density of target electrons.
The value of $L_{ultra}$ reveals a weak dependence on target and
projectile parameters.

In our model the expressions (\ref{LS}) and (\ref{LS_ultra}),
valid for moderately relativistic and ultra-relativistic energies,
respectively, are employed.  In the intermediate energy region we
interpolate between the two approaches using a {\em mix-and-match}
procedure.

\subsection{Low-Energy Corrections}

\subsubsection{Barkas correction}

The Barkas effect, associated with a $z^3$ correction to the
stopping power, is well pronounced at low energies. For
example, for a 2-MeV proton in gold the effect is responsible for about 8\% 
of ionization energy loss~\cite{ICRU37}.  
The correction is due to target polarization effects for low-energy distant collisions 
and can be accounted for by the following expression:
\begin{equation}
\label{Barkas}
   {L_0}(\beta) + {\delta}/2 \rightarrow \left ({L_0}(\beta) + {\delta}/2 \right ) \, \left (1 + 2 \frac{z}{\sqrt{Z}} F(V) \right )
\end{equation}
where $V = \beta \gamma / \alpha \sqrt{Z}$\,.  The function $F(V)$ is a ratio
of two integrals within a Thomas-Fermi model of the atom.
In our model we follow the tabulations for the function from Refs.~\cite{LBNL, Barkas}.

\subsubsection{Shell corrections}

The original Bethe theory is valid when the velocity of the
projectile is much higher than that of electrons in target atoms.
Shell corrections should be taken into account at lower projectile
velocities.  The total shell correction can be presented in the
following form~\cite{ICRU37, ICRU49}:
\begin{equation}
\label{shell}
   {\Delta}L_{shell} = -\frac{C}{Z}
\end{equation}
where $C$ is equal to $C_K + C_L + ...$ and thus takes into
account the contributions from different atomic shells.  For $C_K$
and $C_L$ we follow the asymptotic expressions and tabulations
from Refs.~\cite{Walske52, Khandelwal68} and \cite{Walske56,
Khandelwal68}, respectively, derived with hydrogen-like wave functions.
For all the other atomic shells, up to a combined $O{-}P$ shell, the
scaling procedures developed by Bichsel~\cite{ICRU49} are
employed.  It is assumed in the scaling that the corrections for the
outer shells have the dependence on the projectile velocity similar to 
that of the outermost shell studied with exact calculations, \emph{i.e.}
$L$ shell in our case. 

\subsubsection{Projectile effective charge}

At low projectile velocities, the effect of electron capture and
loss due to interactions with target atoms should be taken into
account as well.  At present, the projectile charge distributions
that cover a more or less noticeable range of ions, targets, and
velocities are not available.  Therefore one can deal with various
empirical and semi-empirical fitting expressions for the average
or, in other words, effective charge, $z_{eff}$.  The effective
charge is to replace the bare projectile charge in all the
relevant expressions.

For protons and other singly charged particles the effective
charge is assumed to be equal to the bare charge down to the lower
energy limit of the model, 100 keV/A. For $\alpha$-particles a
special fit by Ziegler \emph{et al.} \cite{Ziegler} independent of
target material is used at all particle energies, $E$.
\begin{equation}
\label{eff_ch_He}
   z_{eff} / 2 = 1 - \exp \left [ -\sum_{i=0}^{5}a_i {\ln}^i (E) \right ]
\end{equation}
where $E$ is in keV per nucleon and the coefficients $a_0$ through
$a_5$ are equal to $0.2865$, $0.1266$, $-0.001429$, $0.02402$,
$-0.01135$, and $0.00175$, respectively.

For all the other ions more elaborate fitting
expressions that include a dependence on target material are
used:
\begin{itemize}
 \item A combination of the expressions (3.38) and (3.39) from Ref.~\cite{Ziegler} below 1 MeV/A;
 \item The procedure by Hubert \emph{et al.}~\cite{Hubert} above 10 MeV/A;
 \item An energy weighted average between the two energies.
\end{itemize}
For some target nuclei, however, it is necessary to adjust the
upper energy limit to get the stopping power curves with better,
without sharp transitions, appearance.

\begin{figure}[hb]
\begin{center}
\resizebox{4.740in}{!}{%
  \includegraphics{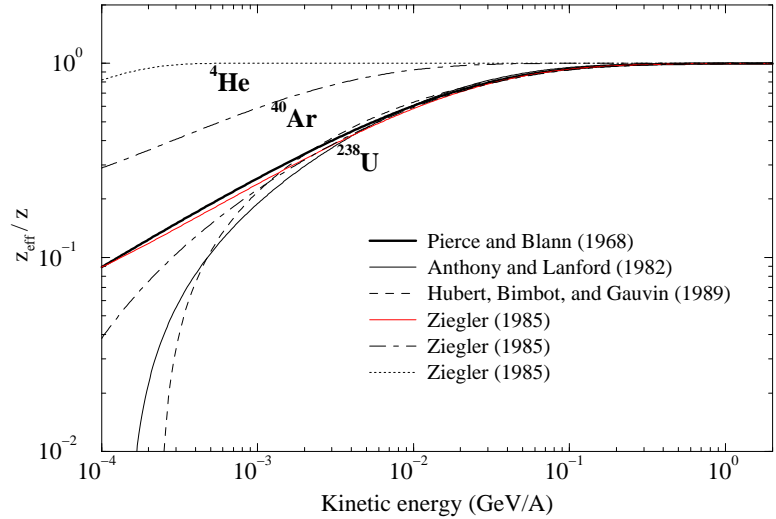} }
\caption{Calculated effective charge of light and heavy ions, $z_{eff}$, in aluminum target relative to
ion unscreened charge, $z$.
\label{eff_charge_in_Al}  }
\end{center}
\end{figure}

Calculated ratios of ion effective charge to bare charge are
presented in Fig.~\ref{eff_charge_in_Al}.  The effect of
neutralization of the bare projectile charge with captured
electrons increases with the target atomic number, being almost
negligible for $\alpha$-particles at energies above a few keV per
nucleon.

\section{Verification}

\subsection{Comparison to experimental data}

\begin{figure}[hb]
\begin{center}
\resizebox{5.975in}{!}{%
  \includegraphics{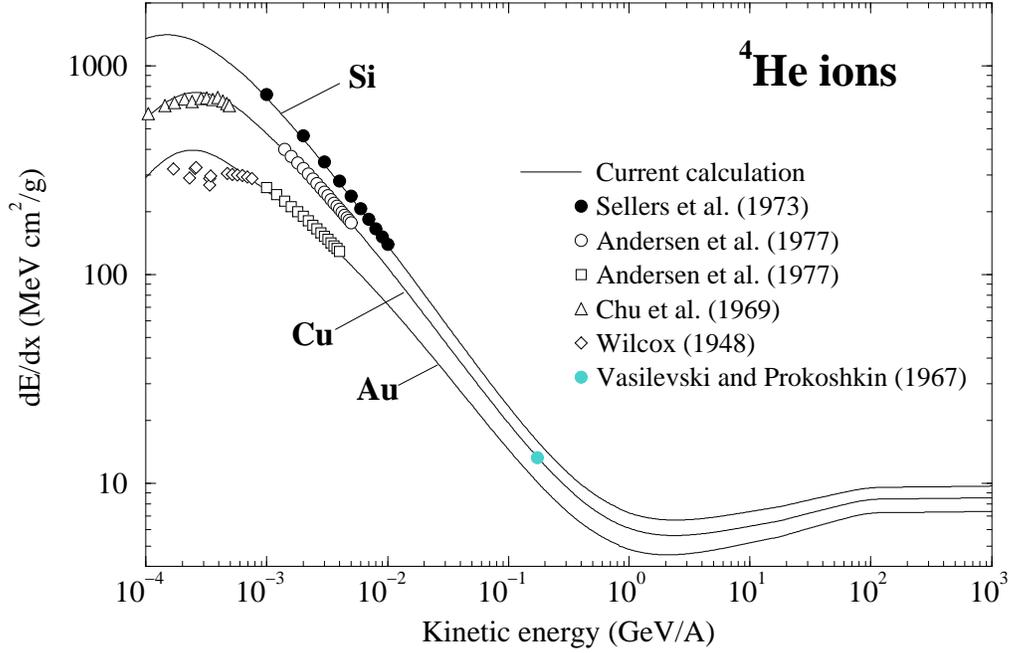} }
\caption{Calculated ionization loss of $\alpha$-particles in various targets \emph{vs.} experimental
data~\cite{alpha_data}. \label{alpha}  }
\end{center}
\end{figure}

Here we compare calculated ionization loss to experimental data for several light and heavy ions.
For $\alpha$-particles at low energies the overall agreement is very good (see Fig.~\ref{alpha}).
The deviations from the Bethe theory due to the above-mentioned corrections, except for the shell
corrections, increase with projectile charge, $z$, at both low and high energies.  Therefore,
the comparisons for super-heavy ions are interesting and important most of all.

\begin{figure}[htb!]
\begin{center}
\resizebox{4.975in}{!}{%
  \includegraphics{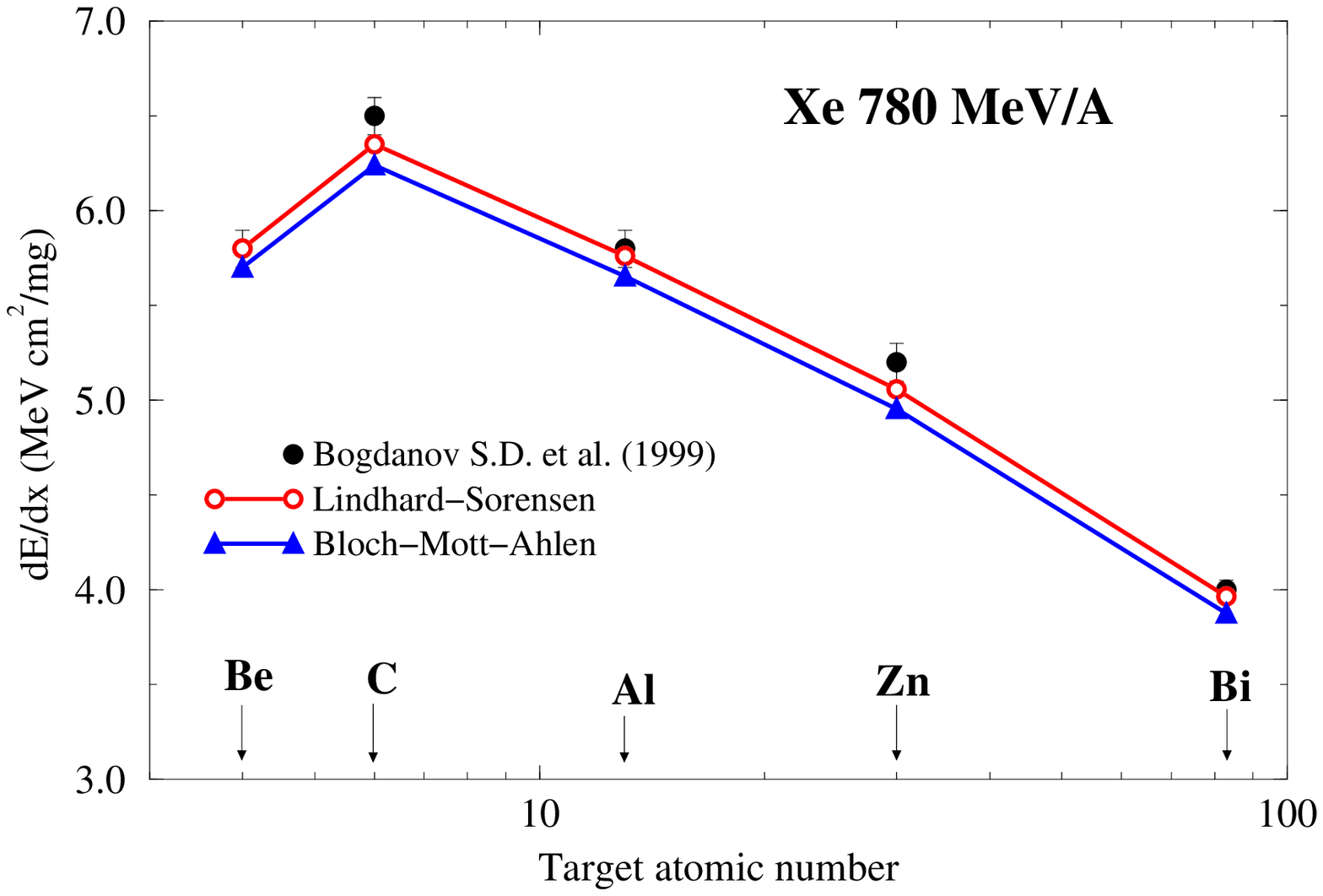} }
\end{center}
 \vspace{-1.7cm}
\begin{center}
\resizebox{4.975in}{!}{%
  \includegraphics{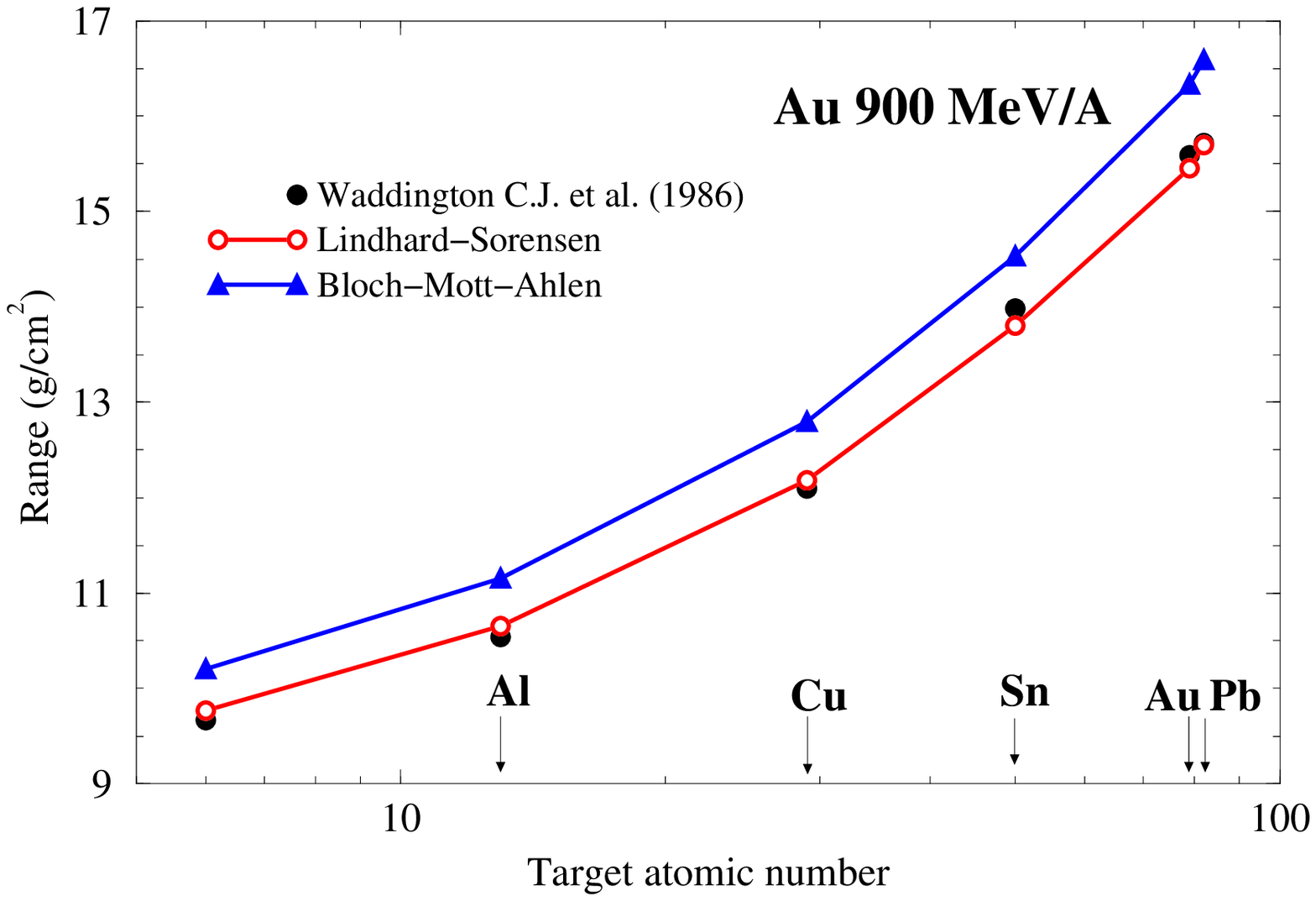} }
\caption{Calculated (lines with symbols) ionization loss and range of relativistic heavy ions 
in various targets \emph{vs.} experimental data (pure symbols)~\cite{Xe_Au_data}.
The corrections to the ionization logarithm, ${\Delta}L$, were calculated following the Lindhard-
S$\o$rensen and Bloch-Mott-Ahlen formalisms (see above). \label{Xe_Au}  }
\end{center}
\end{figure}

At relativistic energies a comparison to experimental data is presented in Fig.~\ref{Xe_Au} 
for a dozen of projectile-target combinations.
One can make the following conclusions from the Figure: (i) the $LS$ correction
in this case provides for an agreement with experimental data within 2\%; (ii) the above-mentioned
combination of relativistic Bloch, Mott, and Ahlen (BMA) corrections gives rise to a systematic 
underestimation of ionization loss (2-3\% for Xe ions) when compared to the $LS$
approach; (iii) the difference between the BMA and $LS$ approaches increases
with projectile charge. This confirms that the Lindhard-S$\o$rensen theory is correctly chosen.

\begin{figure}[htb!]
\begin{center}
\resizebox{4.975in}{!}{%
  \includegraphics{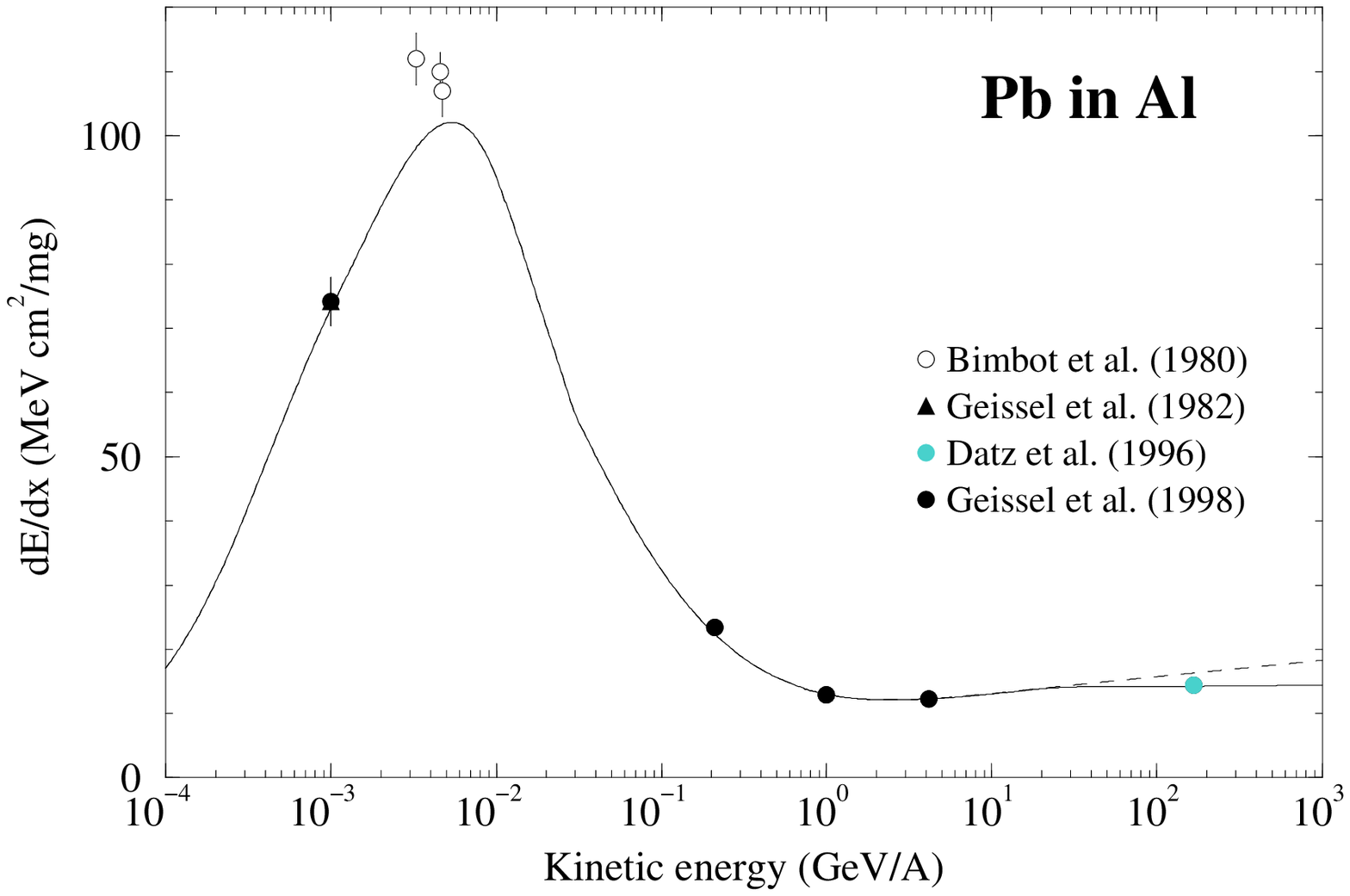} }
\end{center}
 \vspace{-1.7cm}
\begin{center}
\resizebox{4.975in}{!}{%
  \includegraphics{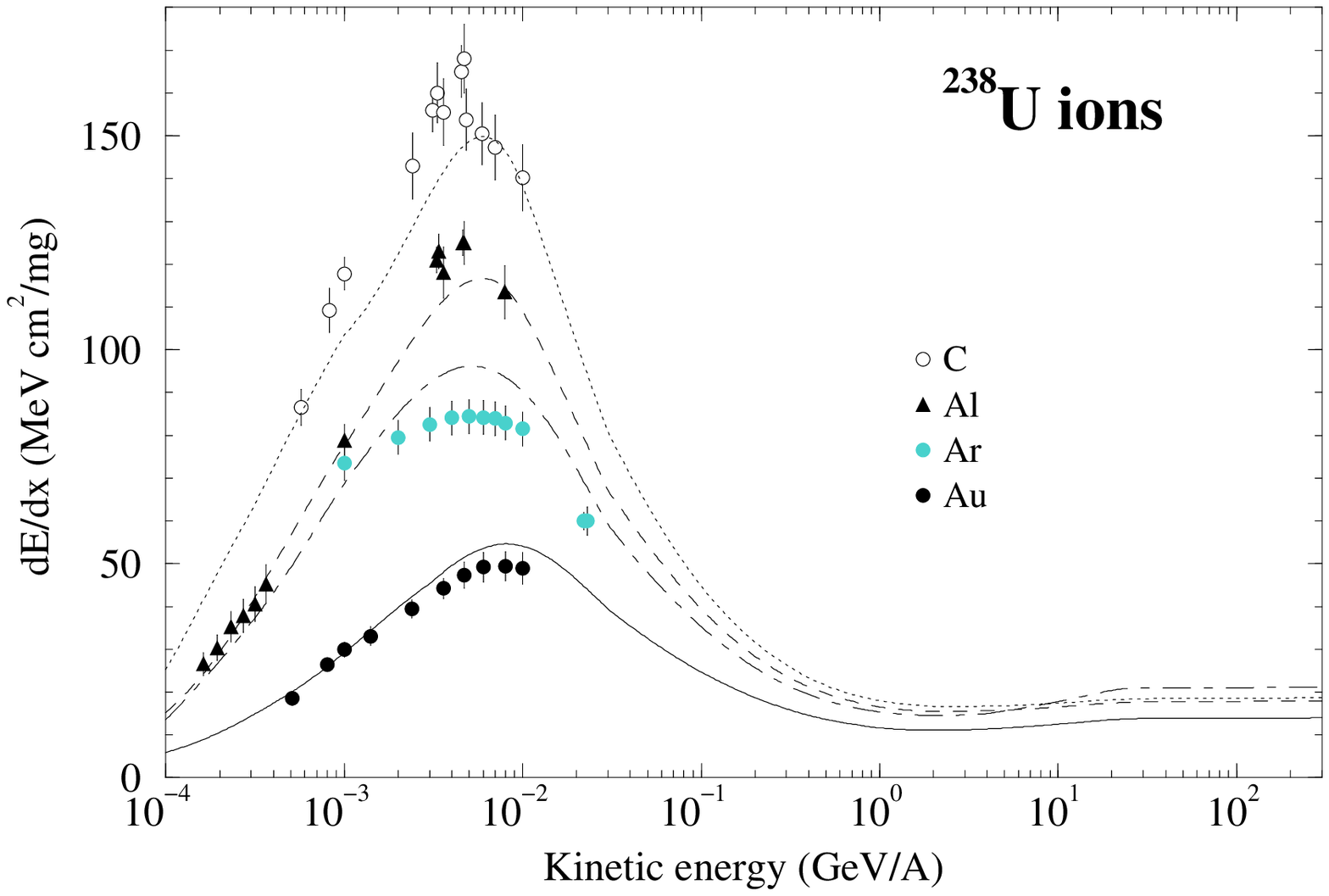} }
\caption{Calculated (lines) ionization loss \emph{vs.} experimental data (symbols) for 
lead ions in aluminum (top) and uranium ions in several targets (bottom)~\cite{U8_data}.
For lead ions the dashed line indicates calculation for pointlike projectiles. \label{Pb_U8}  }
\end{center}
\end{figure}

A comparison to experimental data for super-heavy ions of lead and uranium is given in Fig.~\ref{Pb_U8}.
One can see that the employed \emph{mix-and-match} procedure provides for a good, within 10\%, agreement
with experiment at low energies. For uranium ions the density effect is well seen at ultra-relativistic
energies---the highest ionization loss is observed for the target of the lowest density, \emph{i.e.}
gaseous argon.  For lead ions at ultra-relativistic energies the effect of finite nuclear size,
that gives rise to a saturation of ionization loss instead of a logarithmic growth characteristic of
a pointlike projectile, is easily recognized.  The experimental data at 160 GeV/u by Datz \emph{et al.}
\cite{U8_data} corresponds to the highest energy achieved when accelerating heavy ions.

\subsection{Comparison to NIST data}

\begin{figure}[htb!]
\begin{center}
\resizebox{4.975in}{!}{%
  \includegraphics{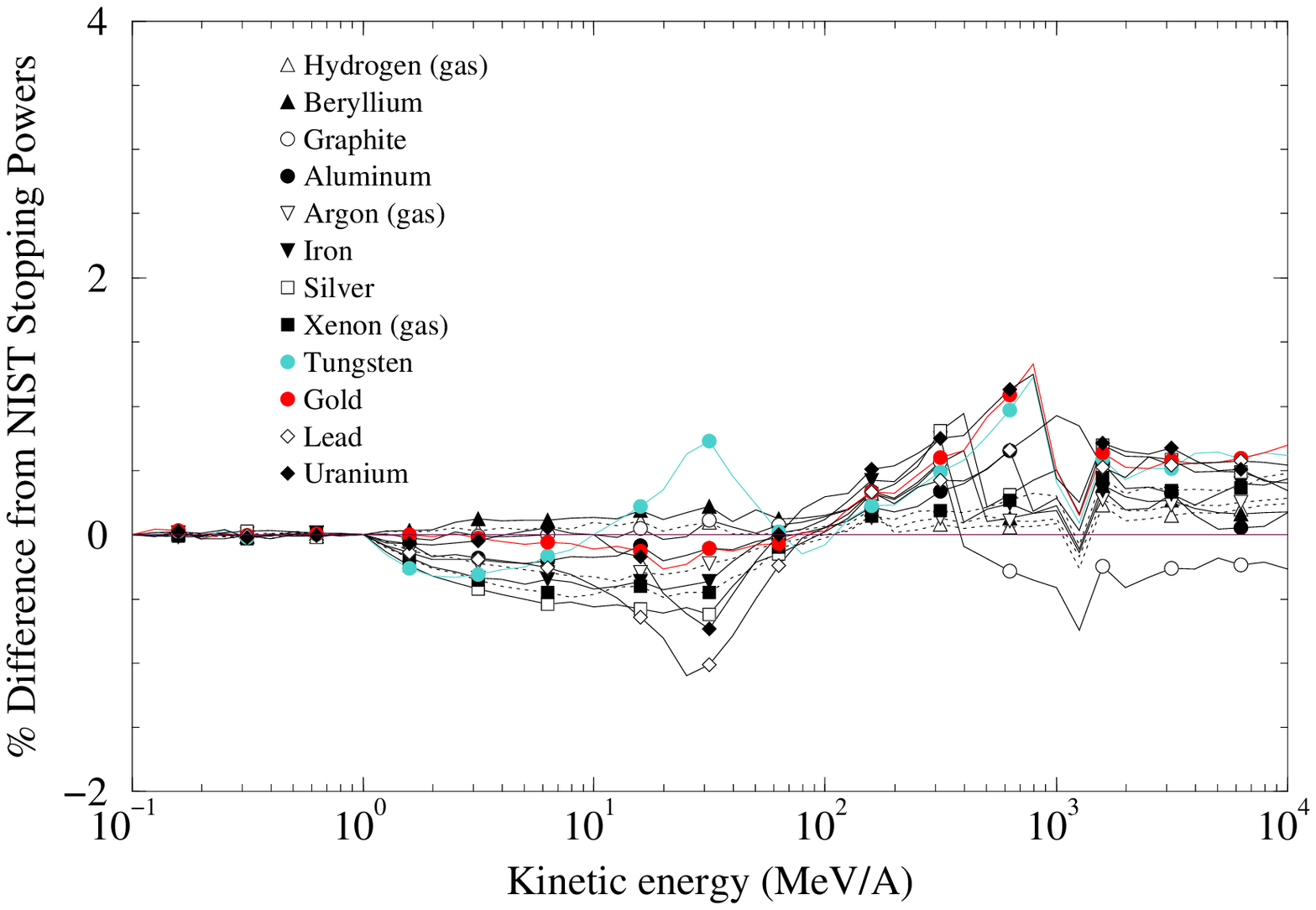} }
\end{center}
  \vspace{-1.7cm}
\begin{center}
\resizebox{4.975in}{!}{%
  \includegraphics{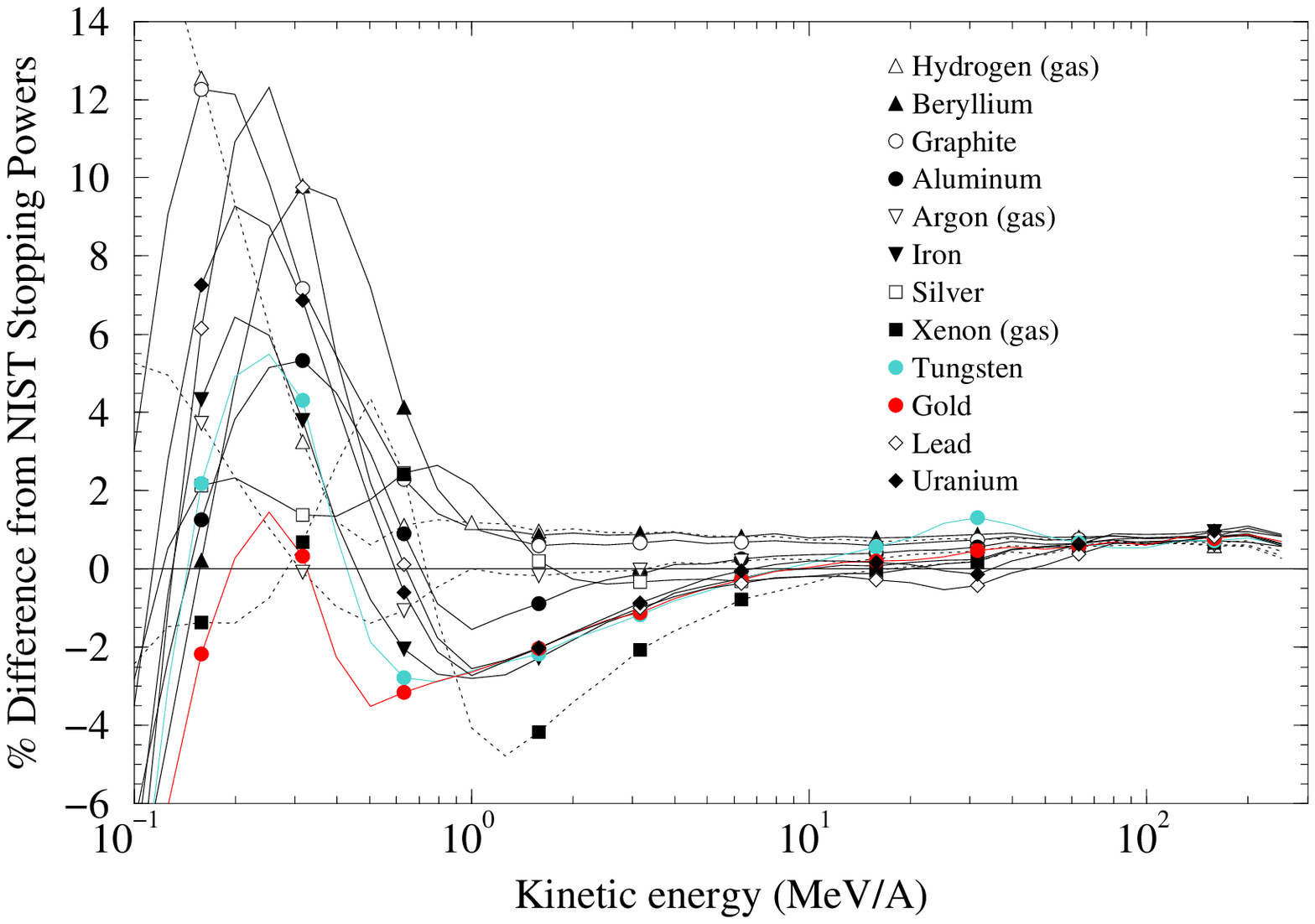} }
\caption{A comparison of \textsc{MARS15} proton (top) and
$\alpha$-particle (bottom) ionization loss in several elements to
NIST data. \label{comparison}  }
\end{center}
\end{figure}

A comparison between the ionization loss calculated within the
framework of the described formalism and the data by
NIST~\cite{NIST} is presented in Fig.~\ref{comparison} for protons
and $\alpha$-particles.
% \begin{figure}[hb]
% \begin{center}
% \resizebox{5.975in}{!}{%
%   \includegraphics{alpha_pct_diff.eps} }
% \caption{A comparison of \textsc{MARS15} $\alpha$-particle stopping powers in several elements to NIST data.
% \label{comparison_alpha}  }
% \end{center}
% \end{figure}
The data of Ref.~\cite{NIST} are given up to $10^4$ MeV and $250$
MeV/A for protons and $\alpha$-particles, respectively.  One can
see that the agreement between the \textsc{MARS15} and NIST
ionization loss is within $1.3$\% for protons in the entire energy
region.  The agreement is somewhat better than that of
MCNP5~\cite{MCNP5} where the difference is about $3$\% for the
energy region from 4 up to $10^4$ MeV, being more than $10$\%
below 4 MeV.

For $\alpha$-particles the biggest difference, about $10$-$15$\%,
is observed below 400 keV/A.  The difference is comparable to the
disagreement between theory and experiment in the energy region.
As far as the tabulated proton data
of Ref.~\cite{NIST} are used below 1 MeV/A in our model, the
differences can be attributed to the description of effective
charge of $\alpha$-particles.  Above 10 MeV per nucleon the
observed difference between the \textsc{MARS15} and NIST
ionization loss is about $1$\%. One can see that approximately a
half of the $1$\% is due to the difference in the description of
the proton ionization loss.

\section{Conclusions}

The various corrections to the Bethe mean ionization loss theory, as implemented in the \textsc{MARS15} 
Monte Carlo code, are described.  The comparisons of calculated ionization loss to the NIST published values
reveal good overall agreement for protons and $\alpha$-particles.  The agreement between the current
model and experimental data is very good up to the super-heavy ions of lead and uranium.

Experimental programs at many accelerator facilities cover wide energy regions.  For example, the Rare Isotope
Accelerator is supposed to be operated at energies from a few keV/A up to hundreds of MeV/A.  To meet 
such practical demands, the developments are underway to validate our model in the 1--100 keV/A region. 

\section{Acknowledgments}

This work was supported by the Universities Research Association,
Inc., under contract DE-AC02-76CH03000 with the U.S. Department of
Energy.

% \appendix

% \section{}

% If necessary, include Appendices by preceeding a section command with
% the appendix command.  The section commands defining different
% appendices should have blank arguments.  The appendices will be
% numbered in upper case alphabetical order. In order to ensure a
% uniform, professional look to the proceedings, do not deviate from
% this format without checking with the organizers first.

\end{document}